\documentclass[12pt,tightenlines]{revtex4}
\usepackage{amssymb}
\usepackage{amsmath}
\usepackage{amsfonts}

\begin{document}

\title{Unification and Emergence in Physics: the Problem of Articulation}

\author{Ian T. Durham}
\email[]{idurham@anselm.edu}
\affiliation{Department of Physics, Saint Anselm College, Manchester, NH 03102}
\date{\today}

\maketitle

\section{What $is$ physics?}
What $is$ physics?  Is there a unifying theme to what we, as physicists, study, whether it be classical or quantum?  How do we effectively convey our ideas $to$ $each$ $other$?  Are there limits to what we may say?  The sciences can very generally be divided into two categories: the life sciences that tend to be taxonomic and statistical and the physical sciences that tend to be procedural.  In some sense, physics is pure, unadulterated `scientific methodology' if you will.  It represents a way of thinking about the world that starts, by one account, \begin{quote}from stable, given facts - observations, measurements in the form of numbers, isolated and purified substances that are part of an unchanging, solid reality.  Logic then compels the assembling (a process carefully controlled by existing theory) of these indisputable pieces of the real world into a theory which literally re-presents that world: a perfect match that, when done correctly, admits no doubt.  The theory is checked by the fact, by the real world, and underwritten by the rigor and purity of the scientific method.  Together, hard fact and reliable method provide [physics] with a unique and powerful tool for self-correction that eliminates (in the long run) all forms of bias and error, yielding a neutral and objective progressive approach to equally neutral and objective final truths. (\cite{Fortun:1998rz}, p. $xii$)\end{quote} Technically the above quote was talking about the sciences in general, but I took the liberty of substituting `physics' for `the sciences' in the quote since it so succinctly describes one very popular view of our discipline.

The reality, of course, is more complicated.  One could say physics represents a way of $doing$ that involves a great deal of trial-and-error: \begin{quote}days, months, and years of what many [physicists] call ``tinkering'' - trying to get a piece of equipment to work properly, interpreting messy data, separating signal from noise, articulating a new theory and explanations for new phenomena.  Any decent [physicist] knows that results and explanations are always open to revision - indeed, must be revised if those results and explanations are to mean or work for anything (\cite{Fortun:1998rz}, p. $xiv$).\end{quote}  Again I took the liberty of substituting `physicist' for `scientist.'  Or, to paraphrase a colleague of mine, we're simply modeling the world of our sensory perceptions as best we can.  We constantly seek out models that are more encompassing, but they are models nonetheless.  This does $not$, however, release us from our obligation to rigor and the strictures of the scientific method.

But there's something more to physics than simply this.  Physics likes to strip out all the extraneous baggage of a problem before reassembling it.  Physics deals with the most fundamental aspects of the universe.  Thus, in that sense, it is the science of simplification.  The best physical theories are both simple and elegant and provide building blocks from which we may re-assemble nature.  In contemplating the explanatory limits of physics, it makes sense to keep this in mind.  But in the process we also must take special care in our use of the language within which we form our ideas.  In the words of Fortun and Bernstein, \begin{quote} Language matters.  Language is essential to reason, and can't be gotten rid of so easily with a few new machines.  Somewhere along the line - no matter how long that line is - every experiment, every mathematical equation, every pure numerical value will have to find its way into words. (\cite{Fortun:1998rz}, p. 43)\end{quote}  And so we begin with that most nebulous of words, `stuff.'

\section{In search of the Holy Grail}
The universe is made up of a lot of interacting `stuff.'  While we, as physicists, may be interested either in the nature of the `stuff' or the nature of the interactions themselves, we can all agree that, without the interactions, the universe would be a very boring place.  In fact, it is often through the interactions that we come to understand the `stuff' since, on a fundamental level, these interactions transfer information in various forms including momentum and energy.  The unification of the four most fundamental interactions via a `theory of everything'  has been a long-standing goal of theoretical physicists \footnote{John Ellis claims to have first introduced this phrase into the serious physics lexicon.  See J. Ellis, Nature, \textbf{415}, 957 (2002) and J. Ellis, Nature, \textbf{323}, 595-598 (1986).}.  For many years string theory was expected to fulfill this dream, or so the story goes.  While still likely to yield very promising results, it appears it may simply be a better refinement of existing quantum field theories, though the jury is still out.  But even if it were to unify the four fundamental interactions, there are numerous other fundamental phenomena on which it is seemingly silent.  Most crucially, it does very little to further our understanding of how the macroscopic world emerges from the microscopic world, nor does it do much to further elucidate the nebulous boundary between the quantum and the classical which may $or$ $may$ $not$ be the same thing as the boundary between `micro' and `macro,' as we shall see.  As physicists we are always seeking broader explanations that include more phenomena under fewer umbrellas, so-to-speak.  The problems these issues present must be successfully addressed by any theory that boasts of being a `theory of everything.'  Indeed the nature of these problems pushes the boundary of what is even possible within physics itself as a discipline, a way of thinking, a way of doing, and a body of knowledge.

In terms of interactions the Standard Model of particle physics and its extensions successfully explains three of the traditional four fundamental of these with gravity the `odd man out.'  Nonetheless, one might concede the possibility that some future extension of the Standard Model could bring gravity into the fold.  In this case, the four fundamental interactions would each be accompanied by a mediator particle - the photon for the electromagnetic, vector bosons ($W^+$, $W^-$, and $Z^0$) for the weak nuclear, gluons for the color, and gravitons for the gravitational.  One could consider the Higgs mechanism and its associated boson as well as being fundamental assuming it is eventually verified.  Higher order microscopic interactions, such as the strong nuclear, possess their own mediator particle (e.g. the meson).  One can theoretically use these as the building blocks for ordinary macroscopic matter \cite{Moore:2003yq} with one glaring exception: the extended structure of the atom.  In addition to two of the fundamental interactions, `building' an atom requires invocation of the Pauli Exclusion Principle (PEP).  PEP may be understood in the context of the Standard Model via the spin-statistics theorem - fields ultimately possess certain commutation properties that manifest themselves, after the action of a field operator, as bosons or fermions, the latter obeying PEP.  In other words, we would find that a certain field $has$ to be commuting (or perhaps anti-commuting) or else we get, in the words of Tony Zee, `a nonvanishing piece of junk' in our mathematics \cite{Zee:2003mz} (p.119).  

How good a $physical$ description is this?  Physically speaking, what is a `nonvanishing piece of junk' exactly?  With its vision of mediating interactions, the Standard Model inherently incorporates special relativity - mediator particles are constrained by the same rules as any other and thus can't carry information about an interaction superluminally.  This is easily visualized.  But what does it mean for a field to possess a $mathematical$ property such as a commutation rule?  Crucially, how does this connect to the reality we experience on a macroscopic level?  I will hold off discussing the former question until a bit later.  As for the latter, as Zee notes \begin{quote}[i]t is sometimes said that because of electromagnetism you do not sink through the floor and because of gravity you do not float to the ceiling, and you would be sinking or floating in total darkness were it not for the weak interaction, which regulates stellar burning.  Without the spin statistics connection, electrons would not obey Pauli exclusion.  Matter would just collapse. (\cite{Zee:2003mz} pp. 119-120)\end{quote}  In other words, constructing the macroscopic world of our senses requires the interaction picture along with a healthy dose of PEP.  

\section{Trouble in paradise}
The macroscopic and classical worlds are usually described in terms of forces, as any veteran of an introductory physics course is likely to tell you.  It is one of the many things engineers rely on when designing systems that we use in our everyday lives.  Forces are easily understood within the context of the interaction picture if they are defined as the rate of momentum transfer between interacting objects - no interaction means no momentum transfer and thus no force \cite{Moore:2003yq}.  (This ignores the question of $why$ we might define them as such.)  In fact, as Zee notes, the fact `[t]hat the exchange of a particle can produce a force was one of the most profound conceptual advances in physics' \cite{Zee:2003mz} (p. 27).  But PEP does not represent an interaction.  In theory, nothing should be transferred between objects associated through PEP meaning we should not be able to associate forces with PEP.  Specifically, within causal quantum field theories, both the commutator (for two-boson fields) and anti-commutator (for two-fermion fields) must vanish for space-like separations \cite{Kaku:1993zl}.  Interactions are causal processes.  Since particles associated through some process via PEP can't be interacting with one another, they must be space-like separated.  One of the axioms of relativistic quantum field theory is thus that any pair of space-like separated observables must commute.

How might the issue of PEP cause much trouble, though, if the Standard Model explains it?  We are already aware of the disconnect between the quantum and classical worlds so couldn't we just write this off as something irrelevant to classical, Newtonian/Einsteinian physics?  Here's why we can't if we work within the interaction picture.  In that case the existence of non-pointlike objects ought to elicit qualms in anyone whose business it is to provide a coherent description of the universe.  The qualms might turn into genuine queasiness for, say, a stellar astrophysicist when he or she attempts to describe a white dwarf star in which the inward pressure of the gravitational interaction is counter-balanced by degeneracy pressure, i.e. PEP (math):  to borrow a bit from physics pedagogy, how can we draw a free-body diagram of a stable chunk of white dwarf star if forces are only associated with interactions?  In fact, given our definition of forces, a stable chunk of white dwarf star violates all three of Newton's laws but not for any relativistic reason.  We seem to have run into a problem in our articulation of certain phenomena while in the process of building up the macroworld from the microworld.

There is, in fact, a deeper problem here that is better elucidated by looking at another seemingly strange phenomenon: entanglement.  Whether entanglement is truly a non-local effect or can be causally explained is a hotly debated topic \footnote{A list of the literature on this subject could fill a book.  To get a rough sense where some leading physicists stand one might take a look at \cite{Arndt:2005db}, though not all the included statements are equally enlightening on the subject.  To quote Artur Ekert from the aforementioned paper, `I like small gadgets, look at this tiny digital camera\ldots where is it\ldots'.}.  Most `interpretations' of quantum mechanics make some attempt at a palatable explanation of this phenomenon, i.e. not many people simply throw up their hands and accept the non-locality of quantum mechanics without question.  We know there is a deep link between the notion of causality (and thus locality) and reversibility, yet we also know that on the microscopic level there are processes that are fully reversible.  Nonetheless, the latter we tend to explain away by assuming that, as we scale up, processes become irreversible as the multiplicity increases \cite{Moore:1997rz, Schroeder:2000zl}.  In other words, we might comfortably assume that these problems of locality and reversibility are confined to the microscopic realm.  As M. Hossein Partovi points out, \begin{quote}The status of the second law of thermodynamics and the emergence of macroscopic irreversibility from time-symmetric dynamics laws have been widely debated since Boltzmann's ground breaking work relating thermodynamic behavior to microscopic dynamics late in the nineteenth century.  Indeed, rarely have so many distinguished physicists written as extensively on a subject while achieving so little consensus \ldots There is nevertheless a slowly growing consensus that the asymmetry observed in macroscopic phenomena originates in the ``initial conditions'' of our cosmic neighborhood, and ultimately that of the whole universe \cite{Partovi:2008rz}.\end{quote}  Indeed, the quantum realm is $usually$ considered to be equivalent to the microscopic realm, though macroscopic quantum phenomena do exist, e.g. flux quantization and the Josephson effect \cite{Leggett:1980rz} (one could argue these are merely macro-manifestations of microscopic phenomena).  In any case, they tie the issue of the emergence of the macroworld from the microworld to the quantum-classical trasition.  Aside from the usual questions of where exactly the quantum-classical transition occurs when scaling up (or down), Partovi himself has introduced a twist to this that raises new questions.  Using models of dilute gases, he constructed a class of macroscopically entangled systems in which heat can flow from the colder to the hotter system.  His results actually bolster Boltzmann's argument that the second law is an `emergent phenomenon that requires a low-entropy cosmological environment, one that can effectively function as an ideal information sink' \cite{Partovi:2008rz}.

What makes Partovi's result so interesting is that it provides additional evidence of macroscopic entanglement while simultaneously suggesting a link between entanglement and the second law of thermodynamics.  Macroscopic entanglement has already been demonstrated in the bulk properties of magnetic materials \cite{Ghosh:2003qd} while a link between entanglement and the second law has been suggested \cite{Durham:2008rc, Durham:2007dn}, but Partovi has managed to unite all of these concepts under a single umbrella.  

All of these results call into question the very meaning of the word `quantum.'  If we hold that `quantum' is synonymous with `microscopic' then entanglement is not a purely quantum phenomenon (and we note that there is a clear difference here between entanglement and mere correlation, where the latter can occur in classical systems).  If we insist that only quantum systems may be entangled, then we're stuck with the realization that `quantum' is $not$ synonymous with `microscopic.'  Either way, we once again are stuck with a macroscopic phenomenon that can't be explained with our neat and tidy interaction picture.  We have also run into another vaguely unsettling discord in our language of description.

\section{Some perspective}
One of the problems here is that, for perhaps bio-psychological reasons, we seem as human beings to prefer axiomatized descriptions of things.  We like rules.  Unfortunately this bumps up against reality in the form of incompleteness: fully axiomatized systems are logically impossible \cite{Godel:1931rz}.  Nevertheless, we need axioms and rules, for without them science becomes pointless and so we do our best with what we have.

Classically, there are two types of such rules in non-relativistic physical theories: $laws$ $of$ $coexistence$ and $laws$ $of$ $succession$.  The former are laws such as the ideal gas law while the latter include the classical laws of motion.  In quantum mechanics these restrictions take the form of selection and superselection rules \cite{Fraassen:1991zl}.  \begin{quote}Symmetries of the model - of which Galileo's relativity is the classical example $par$ $excellence$ - are `deeper' because they tell us something beforehand about what the laws of coexistence and succession can look like.  It is in the twentieth century's quantum theory that symmetry, coexistence, and succession became most elegantly joined and most intricately connected.  (\cite{Fraassen:1991zl}, p.29)\end{quote}  The commutators and anti-commutators we discussed above generally serve as selection rules.  If a given operator commutes with $all$ observables it is a $super$selection operator and we have a superselection rule.  Regarding quantum mechanics, \begin{quote}\ldots there is a conviction that [it] as originally formulated cannot apply universally.  Two ways to say that in contemporary terminology are (a) that the principle of superposition (i.e., for any two pure states $x$ and $y$, there is a pure state $ax+by$) does not have universal validity, or (b) that quantum mechanics must be amended by allowing for superselection rules.  (\cite{Fraassen:1991zl}, p.261)\end{quote}  This leads back to the question of the synonymity of `quantum' and `micro.'  The dividing line between `micro' and `macro' is rather vaguely defined in terms of our human reference frame but humans `have no privileged status in fundamental physics' (\cite{Fraassen:1991zl}, p. 264).  Nonetheless, macroscopic objects are quite clearly more physically complex than microscopic ones.  Van Fraassen's suggestion for dealing with this problem is to wonder whether observables that may be $distinguished$ (and this is an important qualifier) at the level of human observation are subject to restrictions that don't normally hold for $all$ observables.  He provides an example by way of the measurement problem.  The important point here is that superselection rules, as we noted above, are quantum analogues of classical laws of succession.  Thus, if we hold that `quantum' is synonymous with `microscopic' they ought not apply to macroscopic phenomena.  

This relation between superselection rules and macroscopic phenomena is actually testable via the Leggett-Garg inequality \cite{Leggett:1980rz, Leggett:1986zl}.  In quantum mechanics this inequality is violated and means that the time evolution of quantum systems cannot be understood classically.  This muddles the applicability of superselection rules a bit since obeyance of this inequality by macrostates indicates their eigenspaces are, indeed, separated by superselection rules which are normally associated with quantum systems.

But all is not lost.  Recently there has been a movement to understand the quantum-classical transition via coarse-graining.  Kofler and Brukner have proposed such a scheme that does not involve violations of the Leggett-Garg inequality \cite{Kofler:2008fv}.  Nonetheless, that still leaves open the question of the synonymity of `quantum' and `micro' (and of `classical' and `macro').  Thus it is not entirely clear whether we are dealing with two problems or just one.  We may have to acknowledge that `macro' emerging from `micro' may not be the same thing as `classical' emerging from `quantum.'  As such, attempts at any sort of unification of ideas in physics must deal with these problems of emergent phenomena.

\section{Is mathematics a language of description?}
Van Fraassen's point on terminology - language - reminds us of the deep connection to mathematics that all physical theories have.  As we saw in our discussion of PEP, we can have a concise $mathematical$ description of an observable phenomenon with only a vaguely satisfying $conceptual$ description to accompany it.  In fact, Moore describes $all$ of quantum mechanics as being similarly unsettling: the infinite `reality' within which we exist is perceived by us through a finite lens of experimental results and observations.  These are usually given a conceptual description that is then, in classical physics, translated into mathematical symbols.  In other words, the mathematics is merely a shorthand for some conceptual description (think Newton's laws, for instance).  Moore claims it is the conceptual layer that is missing in quantum physics - we have mathematics that match experimental results but no $consistent$ way to conceptually explain the phenomena \cite{Moore:2003rz}.  In paraphrasing Moore, I added the word `consistent' since we clearly $do$ have plenty of conceptual descriptions that we call `interpretations' of quantum mechanics.  Despite one or two of these interpretations dominating the rhetoric at various periods in the past eight decades (e.g. the Copenhagen interpretation - or non-interpretation; see \cite{Beller:1999zl}), there is enough disagreement that new interpretations pop up regularly.  No other area of physics, with the possible exception of the second law of thermodynamics, elicits as much speculation.  One of the most popular of these `interpretations' (if one can call it that) that includes many an experimentalist among its proponents is the almost-serious `shut-up-and-calculate' method (I hesitate to call it an `interpretation' per say - rather it is an $anti$-interpretation or `method').  In a sense, this method assumes that the mathematics do a sufficient job of `interpreting' our experimental results.  But do they?

There are actually two questions that should be asked here.  The first is whether mathematics may serve in an interpretive role, i.e. do we really need a conceptual description?  The second is whether our conceptual descriptions are really just disguised mathematics to begin with.  To some extent this gets at the heart of the nature of mathematics.  Is mathematics discovered or constructed?  Answering this question - if that is even possible - would assist in answering whether or not mathematics may serve in an interpretive role.  If mathematics is `discovered' then perhaps it $is$ sufficient as an interpretive device.  But if it is constructed, then it may simply represent the projection of our perceptions on the physical world.  As Baker puts it, \begin{quote}There is a well-known, ongoing debate concerning the proper ontology for mathematics, between the platonists on one side and nominalists on the other, that arises from the Quine-Putnam indispensability argument.  In its basic form, this argument proceeds from the claim that mathematical objects play an indispensible role in our best scientific theories to the conclusion that we therefore ought to rationally believe in the existence of these mathematical objects.  In other words, scientific realists ought to also be mathematical platonists. (\cite{Baker:2009fv}, p. 612)\end{quote}  Moore is seemingly a mathematical platonist as, perhaps, are advocates of `shut-up-and-calculate.'

Related to this is the question of whether or not our cherished conceptual notions are really conceptual at all.  We are in the habit of not questioning the conceptual aspects of classical, Newtonian physics.  But, for example, can we really conceptually describe energy?  What is energy anyway?  As Feynman says about its conservation, \begin{quote}That is a most abstract idea, because it is a mathematical principle; it says that there is a numerical quantity which does not change when something happens.  It is not a description of a mechanism, or anything concrete; it is just a strange fact that we can calculate some number and when we finish watching nature go through her tricks and calculate the number again, it is the same. (\cite{Feynman:1963lq}, p. 4-1)\end{quote}  Recall the admonishment of Fortun and Bernstein: `\ldots every pure numerical value will have to find its way into words' (\cite{Fortun:1998rz}, p. 43).  We call this particular numerical value `energy.'  But where does that get us?  As physicists, should we really be concerned with the philosophical questions surrounding the nature of mathematics and reality?  I answer an emphatic `yes' to that question.  In attempting to develop a unified description - mathematical, conceptual, or both - of the physical world that satisfactorily explains emergent phenomena, we will inevitably have to confront some of these paradoxical problems.  It is thus imperative that we tackle these issues of articulation head on and, in the process, realize that what is ultimately possible in physics may depend on what is ultimately possible in mathematics.  On a practical level we may need to ask ourselves if mathematics is a sufficient language of description.

\section{Michelson's Legacy}
What I have presented here might simultaneously give us brain cramps and a sigh of relief.  The latter is because we can be assured that our jobs as physicists are safe: there are still plenty of problems left to solve.  Recall Albert Michelson's famous proclamation in 1894 at the opening of the Ryerson Laboratory at the University of Chicago: \begin{quote}The more important fundamental laws and facts of physical science have all been discovered, and these are now so firmly established that the possibility of their ever being supplanted in consequence of new discoveries is exceedingly remote \ldots Our future discoveries must be looked for in the sixth place of decimals.\end{quote}  Luckily (for theorists anyway) we are no closer to this dire prediction than we were in 1894.  A true `theory of everything' is not likely in the near future and any claimant to that dubious title would need to address the aforementioned problems that serve as the potential source of our brain cramps.

Nevertheless, unification is a laudable goal.  If understood in its more general sense - as our desire to find broader explanations to `unite more phenomena under fewer umbrellas' - then any unifying theory must address the issue of how the macroscopic world in which our sensory perceptions reside emerges from the microscopic world.  In order to do that, a unified theory would need to tackle the thorny issues I have raised here.  It would need to incorporate a consistent description of forces and force-like behavior.  It would need to address the question of what it means for something to be called `quantum' and whether or not that is synonymous with `microscopic.'  It would need to better articulate the wholeness of physics.  Whether any of this is possible is an open question.  But keeping these ideas in mind while pursuing unifying themes within physics is imperative, particularly if we want physics to possess as logical and consistent a framework as possible, but also from the standpoint of how we communicate (articulate) with each other as well as with the general public.  As Fortun and Bernstein said, `language matters.'

\begin{acknowledgements}
The main idea for this essay was formulated during a discussion in a pub in Amherst, Massachusetts during the first workshop of the Anacapa Society (for theoretical and computational physicists at primarily undergraduate institutions).  I thus wish to acknowledge the participants of that lively discussion who unwittingly helped me craft my argument.  Specifically, I thank Mary Alberg, Gabriel Cwilich, Will Kennerly, Don Spector, Parker Troischt, Martin Veillette, and Brian Wecht.  Good discussions about physics and good beer are an excellent combination.  I also thank Mike Fortun, Herb Bernstein, and the National Science Foundation for providing the Anacapa Society with the resources to allow some of us to spend time studying their book.  It served as a great source of inspiration for this essay as well.  The only drawback was that it didn't come with beer.
\end{acknowledgements}

\bibliographystyle{apsrev}
\bibliography{FQXiBib.bib}
\end{document}